# Implementing Grover Algorithm on Quantum Chip Architecture Optimized with QGHNN for Fidelity and Entanglement Preservation

Ahmad Salmanogli[1,2] and Hesam Zandi[2,3,4]

[1]Department of Electrical and Electronics, Engineering Faculty, Ankara Yildirim Beyazit University, Turkey
[2]Iranian Quantum Technologies Research Center (IQTEC), Tehran, Iran
[3]Faculty of Electrical Engineering, K. N. Toosi University of Technology, Tehran, Iran
[4]Electronic Materials Laboratory, K. N. Toosi University of Technology, Tehran, Iran

**Abstract:**
This work presents a superconducting quantum-chip architecture designed to simultaneously preserve strong entanglement and high readout fidelity—two performance metrics that are often in competition in scalable quantum hardware. In parallel, the chip architecture is optimized using an AI-driven QGHNN to enhance entanglement and state-separation fidelity, and also increase the Grover's algorithm accuracy. In conventional superconducting circuits, strong qubit–qubit coupling can improve Grover algorithmic accuracy and entanglement but it degrades measurement fidelity. To address these challenges, we propose a hybrid nine-qubit architecture composed of interior and exterior transmon qubits interconnected via a flux-tunable qubit and a distributed resonator network. The interior qubits, together with the tunable qubit, form a controllable entanglement core and along with, are exploited to enhance the Grover algorithm performance, while the exterior qubits operate deep in the dispersive regime to support high-fidelity readout. The full system Hamiltonian incorporates all relevant direct and mediated interactions among interior, exterior, and resonator modes. By numerically solving the Hamiltonian in conjunction with the Lindblad master equation, the system dynamics, spectroscopic behavior, and state-separation fidelity are analyzed. The results demonstrate robust entanglement supported by an engineered avoided-crossing region, while maintaining state-separation fidelity close to 0.995 under realistic noise conditions. Furthermore, QGHNN-based optimization significantly reshapes the device Hamiltonian, yielding an enhanced state-separation fidelity close to 0.9988. Finally, the Grover algorithm implemented on a selected two-qubit subsystem (one interior qubit and the tunable qubit) achieves an accuracy comparable to that of the IBM Q5 Tenerife processor, validating the effectiveness of the proposed architecture optimized with QGHNN.

**Introduction:**
Quantum chip design lies at the heart of advancing quantum technologies, integrating physical qubit realization, control electronics, and error-correction architectures to enable scalable and reliable quantum computation. Typically, quantum chips are built using superconducting circuits, silicon spin qubits, quantum dots, trapped ions, or photonic platforms [1,2]. Key design challenges include maximizing coherence time, Purcel rate reducing, minimizing error rates, achieving precise control of quantum gates, interconnect complexity (especially wiring between qubits and control electronics), heat dissipation at cryogenic temperatures, and ensuring material and device uniformity [1, 5-7]. To analyze and address these, quantum chip design leverages specialized Electronic Design Automation (EDA) tools adapted for quantum-specific constraints such as the mapping of logical qubits to physical devices, layout optimization for low noise, hybrid quantum chip-CMOS architectures, and fabrication process compatibility [2-7]. The applications of quantum chips span quantum simulation of chemical systems and materials, quantum cryptography for secure communications, optimization, quantum metrology, quantum networking, and eventually fault-tolerant universal quantum computers [1,3]. As integration improves—combining qubit arrays with cryogenic control circuitry and interconnects—the gap between laboratory demonstrations and real-world deployment narrows, promising transformative impacts across computing, materials science, and communication technologies [2,4].

Based on an extensive review of the literature [8-14], it is evident that several fundamental quantum properties strongly influence quantum chip design—most notably entanglement, fidelity, and also algorithm accuracy (in this work Grover Algorithm accuracy). Grover's search algorithm is one of the most prominent quantum algorithms [30, 31], offering a quadratic speedup over classical approaches for searching an unstructured database of N elements, requiring only $O(\sqrt{N})$ operations. Owing to its simple circuit structure and strong reliance on core quantum resources—namely superposition, entanglement, and phase-controlled interference—Grover's algorithm has become a standard benchmark for evaluating quantum hardware performance. In particular, the algorithm is highly sensitive to gate fidelity, qubit–qubit coupling strength, decoherence, and readout accuracy, making it an effective probe of a quantum chip's operational quality. For this reason, most superconducting and solid-state quantum processors routinely implement Grover's algorithm to assess their capability to coherently manipulate quantum states and reliably amplify a marked target state under realistic noise conditions [30]. The mentioned characteristics (entanglement, fidelity, and Grover algorithm accuracy) are crucial, as they directly determine the reliability of quantum information processing and the scalability of multi-qubit architectures [1,2]. This raises a central research question: Can a quantum chip be engineered to maintain both strong entanglement and high fidelity simultaneously and in the same way introduce a high Grover algorithm accuracy? In other words, does an inherent trade-off exist between these essential properties? Addressing this question forms the core motivation of the present study. The underlying concept of this work originates from the detailed analysis of two distinct categories of previous research, each exploring different aspects of qubit coupling and performance optimization. In the first category [8, 9], the authors investigated a quantum system comprising two qubits coupled via a resonator and conducted spectroscopic measurements. Their results revealed an avoided level crossing, which signifies the strength of entanglement between the two qubits under a specific external flux bias. The size of this avoided crossing gap is determined by the qubit–qubit transverse coupling coefficient $J_c$. As $J_c$ approaches zero, the gap diminishes, indicating the suppression of entanglement between the qubits. In contrast, studies in the second category [10-14] focus on tunable qubit architectures designed to suppress next-nearest-neighbor (NNN) couplings in order to enhance operational fidelity. In a circuit quantum electrodynamics (cQED) system with multiple transmon qubits coupled through resonators, interactions among higher energy levels give rise to cross-Kerr terms that can be expressed as $\chi a_i^+ a_i a_j^+ a_j$, where $a_i$ ($a_i^+$) denote the annihilation (creation) operators for qubit modes, and $\chi$ affected by the coupling factor between the qubits and bus resonator. This static ZZ (qubit-qubit) crosstalk leads to fidelity performance degradation, particularly when $\chi$ becomes comparable to the qubit decoherence rate, in contrast enhance the Grover algorithm accuracy [30, 31]. Such crosstalk limits the fidelity of XX-type parity measurements used in quantum error-correction protocols and reduces the lifetime of logical qubits involving XX-type stabilizers. Both theoretical and experimental investigations have demonstrated that ZZ crosstalk has become a major constraint on gate fidelity as coherence times continue to improve in state-of-the-art superconducting devices [9-13]. Recently, a simple and broadly applicable scheme has been proposed that employs a tunable coupler to realize high-fidelity two-qubit gates. This approach utilizes a generic three-body system with exchange-type interactions, where a central coupler element mediates a frequency-tunable virtual exchange between two qubits. The coupler exhibits a critical bias point at which the mediated exchange precisely cancels the direct qubit–qubit interaction, effectively turning off the net coupling. Therefore, the ZZ crosstalk which is an original factor to enhance the Grover algorithm accuracy, or maybe a factor to change the entanglement behavior, should be nullified to enhance the fidelity. Thus, comparing the two categories reveals a fundamental trade-off: direct qubit–qubit coupling enhances Grover algorithm accuracy and entanglement but simultaneously degrades gate fidelity and introduces quantum errors. As we know the entanglement (quantum correlation) has been utilized in many quantum systems and applications [15-19]. To address this issue, we propose and analyze a new quantum chip architecture that mitigates this trade-off. The proposed and optimized quantum chip with Quantum Graph Hamiltonian Neural Network (QGHNN) [22–26] enables both high fidelity, entanglement, and high accuracy for Grover algorithm, thereby providing a balanced and scalable solution for next-generation superconducting quantum processors. In the routine of the design, we observed that manually optimizing the parameters of a complex multi-qubit superconducting circuit is extremely time-consuming and quickly becomes infeasible as the system dimensionality grows. This motivated the use of advanced neural-network–based optimization methods capable of handling high-dimensional quantum design spaces. Among these approaches, the QGHNN [22–26] stands out as a powerful, physics-informed framework in which the neural network learns the structure of the system Hamiltonian and efficiently optimizes its critical variables. QGHNN treats the quantum device as a graph, where qubits, resonators, and couplers form nodes and edges, and embeds the Hamiltonian parameters into a learnable graph representation. By training on parameters listed–fidelity datasets generated from the full master-equation simulations, the QGHNN constructs a smooth model that approximates the system's dynamical response without explicitly solving the Lindblad equation during inference. This dramatically reduces computational cost and enables rapid exploration of the parameter landscape. Moreover, because the QGHNN encodes the physical connectivity of the circuit, it leads to optimize the results that remain physically interpretable and directly transferable to hardware design. Finally, the performance of the optimized quantum chip design is assessed using Grover's search algorithm [30,31], to show how the trade-off between the high fidelity, entanglement, and algorithm accuracy is solved.

**Theoretical background:**
*Quantum system definition:*
The proposed quantum chip integrates a hybrid coupling scheme designed to simultaneously optimize entanglement and fidelity. The quantum chip is schematically shown in Fig. 1. As clearly shown, it integrates nine qubits arranged into two distinct groups: the interior group (entanglement core) and the exterior group (readout qubits). The interior group consists of five qubits strongly coupled through shared resonators, forming the entanglement core. These qubits are not directly used for readout but instead play a central role in generating and maintaining nonclassical correlations within the cell. Importantly, one of the interior qubits is designed to be tunable, allowing dynamic enhancement of nonclassicality in the circuit. Unlike common tunable-qubit applications, where tunability is exploited for frequency matching or readout fidelity, here it is deliberately employed to boost quantum correlations and entanglement strength. In this group, the resonator frequencies are deliberately matched to the qubits (e.g., $\omega_{q1A}=\omega_{11}$, $\omega_{q2A}=\omega_{22}$, …), ensuring that the entanglement core operates close to the tunable qubit frequency ($\omega_c$). Additionally, the tunable resonator frequency $\omega_c$ can be swept within a designated band to enhance interaction flexibility. On the other hand, the exterior group consists of four qubits coupled to interior qubits via four dedicated resonators ($\omega_1$, $\omega_2$, $\omega_3$, $\omega_4$). These resonators serve as the interface between the exterior and interior qubits. Their frequencies are selected to satisfy the dispersive readout condition, $|\omega_{qi}-\omega_i|\gg g_i$, ensuring that measurement occurs without destroying the qubit states. This ensures qubit–resonator detuning large enough to suppress direct energy exchange while enabling robust state-dependent frequency shifts. Consequently, this flexible topology allows the chip to operate in two distinct regimes: (i) the entanglement-enhanced mode, where strong inter-qubit coupling yields avoided crossings observable in spectroscopy, and (ii) the high-fidelity mode, where the exterior qubits along with interiors is designed in such a way to enhance fidelity. Such a reconfigurable design supports scalable implementations of error-corrected superconducting qubit systems and provides a practical route to achieving both coherence-preserving operations and tunable entanglement on the same hardware platform.

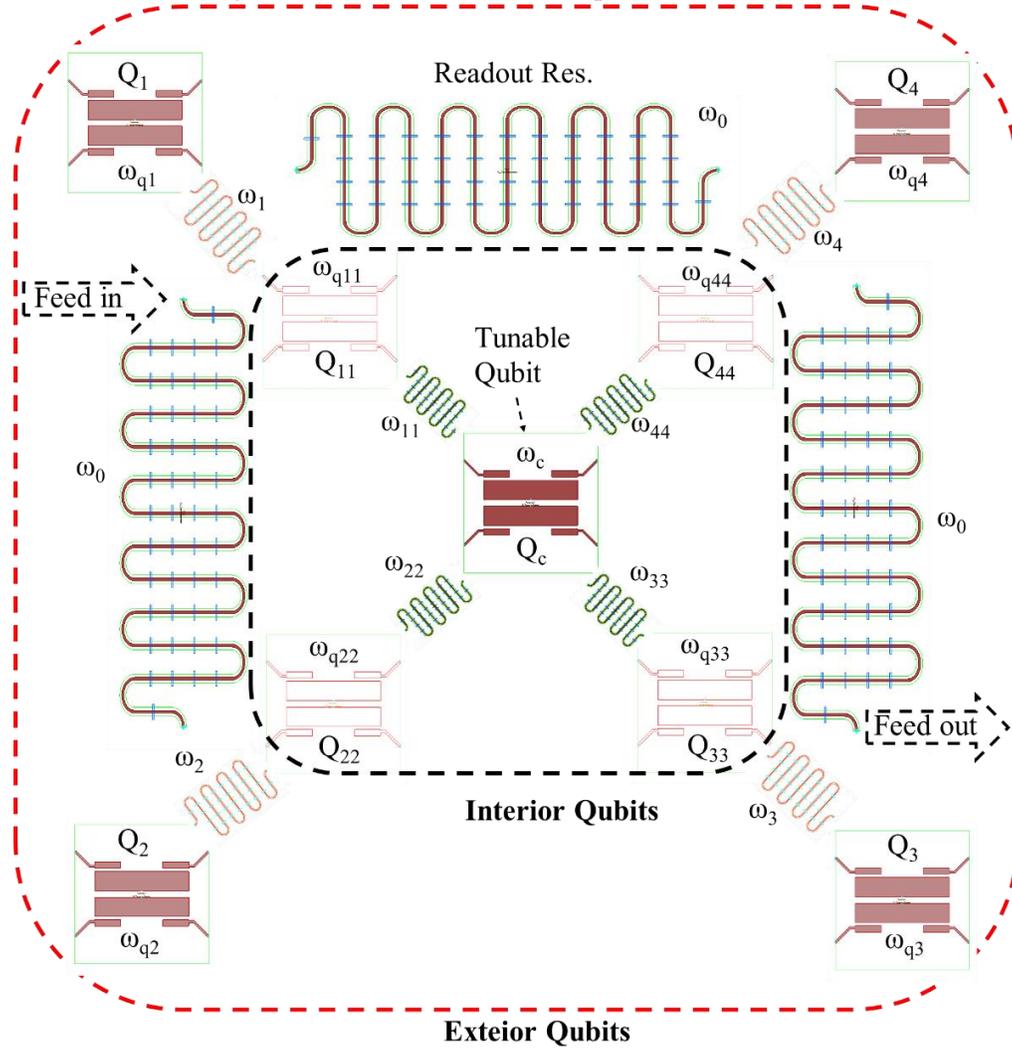

Fig. 1 Schematic layout of the proposed multilayer superconducting quantum chip architecture. The quantum system consists of four exterior qubits ($Q_1$-$Q_4$) and four interior qubits ($Q_{11}$, $Q_{22}$, $Q_{33}$, $Q_{44}$), all capacitively coupled through a central tunable qubit ($Q_c$). The readout resonators ($\omega_0$) are positioned along the perimeter and linked to feedline ports for input and output coupling. Each qubit has its transition frequency ($\omega_{qi}$). The red dashed boundary outlines the exterior qubit region, while the black dashed boundary encloses the interior interaction zone.

*Total Hamiltonian:*

The total Hamiltonian of the quantum circuit depicted in Fig. 1 contains every entity energy and also all interactions among the components is presented in Eq. 1 [9]. The first term is the interior qubits Hamiltonian in which $b_\lambda$, $b_\lambda^+$ are the annihilation/creation operators for the $\lambda^{th}$ interior transmon mode (treated as weakly anharmonic oscillators), and $\omega_\lambda$ is the fundamental transition frequency of interior qubit $\lambda$ (units rad/s). In this term, $\alpha_\lambda$ is the transmon qubit anharmonicity (often called the Kerr term); it makes the level spacing non-uniform so the device functions effectively as a qubit. Finally, the quartic term $(b_\lambda^+ b_\lambda^+ b_\lambda b_\lambda)$ is the lowest-order nonlinearity that produces the qubit two-level behavior.

$$H_{\text{total}} = \underbrace{\sum_{\lambda=1}^{4}[\hbar\omega_\lambda\, b_\lambda^\dagger b_\lambda - \frac{\hbar\alpha_\lambda}{2}(b_\lambda^\dagger b_\lambda^\dagger b_\lambda b_\lambda)]}_{\text{interior qubits}} + \underbrace{\sum_{j=1}^{4}[\hbar\omega_j\, d_j^\dagger d_j - \frac{\hbar\beta_j}{2}(d_j^\dagger d_j^\dagger d_j d_j)]}_{\text{exterior qubits}}$$

$$+ \underbrace{\sum_{k=1}^{4}\hbar\omega_{c,k}\, c_k^\dagger c_k}_{\text{intermediate / coupling resonators}} + \underbrace{\hbar\omega_r\, a^\dagger a}_{\text{readout resonator}} + \underbrace{\hbar\omega_t(\phi)\, t^\dagger t - \frac{\hbar\alpha_t}{2}(t^\dagger t^\dagger t t)}_{\text{tunable qubit}}$$

$$+ \underbrace{\sum_{\lambda=1}^{4}\hbar g_{\lambda b}(c_\lambda^\dagger b_\lambda + c_\lambda b_\lambda^\dagger)}_{\substack{\text{coupling:}\\ \text{intermediate resonators}\leftrightarrow\text{interior qubits}}} + \underbrace{\sum_{j=1}^{4}\hbar g_{jr}(d_j^\dagger a + d_j a^\dagger)}_{\substack{\text{coupling:}\\ \text{exterior qubits}\leftrightarrow\text{readout}}} \quad (1)$$

$$+ \underbrace{\sum_{\lambda=1}^{4}\hbar J_{\lambda t}(b_\lambda^\dagger t + b_\lambda t^\dagger)}_{\substack{\text{effective (SW)}\\ \text{interior qubits}\leftrightarrow\text{tunable qubit}}} + \underbrace{\sum_{\lambda=1}^{4}\hbar J_{\lambda j}(b_\lambda^\dagger d_j + b_\lambda d_j^\dagger)}_{\substack{\text{effective (SW)}\\ \text{interior qubits}\leftrightarrow\text{exterior qubits}}}$$

$$+ H_{\text{drive}}$$

The second term is the exterior qubits (local) Hamiltonian in which $d_j$, $d_j^+$ are the ladder operators for the $j^{th}$ exterior transmon and $\omega_j$, $\beta_j$ are the exterior qubit frequency and anharmonicity, respectively. These qubits in layout can couple to the interior network via the intermediate resonators. The third term is the coupling / intermediate resonators, where $c_k$ are bosonic operators for the intermediate resonators (one resonator per interior–exterior pair in the model). Each resonator frequency is $\omega_{c,k}$. These resonators mediate the interactions between qubits. The fourth term is the readout resonator a, $a^+$ are the readout resonator operators with frequency $\omega_r$. This resonator is driven and is used to obtain the output field <a> that is integrated to produce the I/Q measurement. In this equation, the fifth term is the tunable transmon Hamiltonian, in which $t$ is the annihilation operator for the flux-tunable transmon qubit. Its frequency $\omega_t(\phi)$ depends on the applied flux $\phi$ (through the Josephson energy $E_J(\phi)$): this is the control knob that moves the tunable qubit into/out of resonance and produces the avoided crossing in spectroscopy and also, $\alpha_t$ is its anharmonicity. The sixth term is the coupling terms between intermediate resonators and interior qubits, where $g_{\lambda b}$ is the coupling between resonator $c_\lambda$ and interior qubit $b_\lambda$. These terms are responsible for coherent excitation exchange between resonators and qubits when near resonance. The seventh term is the readout–exterior qubits coupling, in which $g_{jr}$ couples exterior qubits $d_j$ to the readout resonator $a$. This enables those qubits to imprint a state-dependent shift on the readout tone. In addition, the eighth term (two successive terms) is the effective exchange couplings (Schrieffer–Wolff mediated terms), in which $J_{\lambda t}$ and $J_{\lambda j}$ are effective (second-order) exchange couplings between modes and appear after eliminating the intermediate resonator degrees of freedom. A commonly used closed form in the dispersive limit is: $J \sim 0.5 g_1 g_2 (1/\Delta_1 + 1/\Delta_2)$, where $g_{1,2}$ are the bare couplings to the intermediate mode and $\Delta_{1,2}$ are the detunings of the two qubit-like modes from that intermediate resonator. Finally, the drive term, where $H_{\text{drive}} = \mathcal{E}(t)a + \mathcal{E}(t)^* a^+$, where $\mathcal{E}(t)$ is the (complex) amplitude of the measurement/probe tone. The readout drive establishes the steady-state amplitude of the resonator, thereby defining the measurable output field $\langle a(t)\rangle$, which is subsequently incorporated into the Lindblad master equation to compute the corresponding I/Q signals. The following section presents the theoretical framework employed to evaluate the separation fidelity by using the total Hamiltonian derived and solving the Lindblad master equation.

*Separate fidelity:*

In this section, we attempt to show that how the total Hamiltonian build maps to the measured IQ signals, and how those signals are turned into a separation fidelity, and how SNR controls and manipulates the fidelity. In a standard open-system language the dynamics of a quantum system are governed by the Lindblad master equation presented as [9, 18]:

$$\dot{\rho}(t) = -\frac{i}{\hbar}[H_{\text{total}}, \rho(t)] + \sum_j D[c_j]\rho(t) \qquad (2)$$

where $D[c]\rho = c\rho c^+ - 0.5 \times \{c^+c, \rho\}$ and $\rho$ is the density matrix and $c$ and $c^+$ *are* collapse operators (readout decay, resonator loss, qubit relaxation, and so on) [18]. To measure I/Q signals, it needs to solve the Lindblad master equations and collect the expectation value of $\langle a(t)\rangle = \text{Tr}[a\rho(t)]$, then integrate this over the measurement window $T_{\text{int}}$ as [9, 20-21]:

$$I = \int_0^{T_{\text{int}}} \text{Re}[\langle a(t)\rangle]\, dt, Q = \int_0^{T_{\text{int}}} \text{Im}[\langle a(t)\rangle]\, dt \qquad (3)$$

This gives the complex mean vector $\mu = (I, Q)$ for each prepared state ($\mu_0$ and $\mu_1$). To produce the distribution of single-shot I/Q outcomes caused by quantum measurement noise and jump events, it is necessary to collect per-trajectory integrated I/Qs across trajectories (i.e., integrate the output of each trajectory separately). However, in this work, we compute mean signals and then adds classical Gaussian noise later — that is a perfectly valid model for added amplifier/classical noise, but it does not automatically include the full quantum single-shot variability. After integration, there are two mean vectors in I/Q plane: $\mu_0$ and $\mu_1 \rightarrow \mathbb{R}^2$. It generates measurement outcomes by adding (classical/assumed) Gaussian noise as I/Q = $\mu + n$, $n \sim N(0, \Sigma = \sigma^2_{IQ})$, where $\sigma_{IQ} \sim \kappa_{SNR} \times |\mu_1 - \mu_0|$. The factor $|\mu_1 - \mu_0|$ determines the signal separation. Then it projects outcomes onto the optimal axis $u = \mu_1 - \mu_0$, and $u^* = u/\|u\|$ shifts by the midpoint $m = (\mu_1 + \mu_0)/2$, and computes scalar projections $X = (I/Q - m) \times u^*$. With the midpoint threshold, classifying $X > 0$ as state 1 and $X \leq 0$ as state 0 is the optimal linear test when the two classes are Gaussians with equal isotropic covariance. If the projected conditional distributions are Gaussians with equal variance $\sigma^2$, the probabilities for two different states are given by:

$$x \mid 0 \sim \mathcal{N}(-\frac{\Delta\mu}{2}, \sigma^2), x \mid 1 \sim \mathcal{N}(+\frac{\Delta\mu}{2}, \sigma^2) \qquad (4)$$

where $\Delta\mu = |\mu_1 - \mu_0|$. With the midpoint threshold, the correct classification probabilities are identical and equal to $\Phi(\Delta\mu/2\sigma)$, where $\Phi$ is the standard normal CDF. Thus, the separation fidelity (average correct classification probability for equal priors) obeys the formula $\Phi(\Delta\mu/2\sigma)$. Using the derived formula, the separate fidelity is calculated; in the next section, we attempt to illustrate the results related.

**Results and Discussions:**

In this section, we address the central question of this study: Can a quantum chip be designed to simultaneously sustain strong entanglement and high fidelity? To answer this, we demonstrate that the proposed architecture preserves robust entanglement while maintaining high readout fidelity under realistic noise conditions. For this reason, some critical numerical simulation results are presented in Fig. 2. Fig.2a presents the variation of the separation fidelity as a function of the signal-to-noise degradation factor, $\kappa_{SNR}$, which scales the effective measurement noise in the I/Q plane. Each data point is extracted from Monte Carlo simulations of the dispersive readout using the full many-body Hamiltonian. The trend demonstrates that the fidelity remains near unity for small $\kappa_{SNR} < 0.18$, confirming that the readout process can reliably clearly distinguish the logical states |0> and |1> when the measurement noise is weak. As $\kappa_{SNR}$ increases, the I/Q distributions of the two qubit states start to overlap, reducing the discrimination efficiency. Beyond $\kappa_{SNR} \sim 0.22$, the fidelity rapidly decreases, highlighting the exponential sensitivity of classification accuracy to SNR. This curve quantitatively reflects the analytic relationship $F \alpha \ 1/\kappa_{SNR}$, linking the measurement fidelity to the SNR through the separation distance of the I/Q clusters. However, the key question is whether the designed quantum circuit can effectively generate and sustain entanglement. For this, Fig.2b depicts the simulated spectroscopy of the quantum system as a function of the applied flux $\phi/\phi_0$ to the tunable qubit and the probe frequency. The color scale represents the normalized transmission or reflection intensity obtained from the system's response to a weak probe. The key spectral feature is the avoided crossing near $\phi/\phi_0 \sim 0.82$, which arises from the flux-tunable transmon interacting coherently with the tunable qubits. This avoided crossing signifies hybridization of their eigenstates, forming entangled dressed states whose splitting magnitude encodes the effective coupling rate $J_{\lambda t}$ derived from the Schrieffer–Wolff transformation. At other flux biases, the tunable qubit is far detuned and the modes remain largely independent, producing well-separated resonance branches. The continuous evolution of the transition frequency with flux demonstrates both the flux dependence of the tunable qubit's Josephson energy and the controllable nature of the qubit–resonator–qubit coupling network. Therefore, Fig. 2b provides direct spectroscopic evidence of the entanglement inherent in the total Hamiltonian and identifies the optimal bias region ($\phi/\phi_0 \sim 0.82$) for nonclassicality creating in the quantum circuit. So far, the results have shown that by controlling the SNR, it is possible to achieve a high separation fidelity, while the proposed quantum circuit also demonstrates a strong capability to generate entanglement. Finally, the analysis focuses on examining how the separation fidelity behaves around the optimal flux bias, where the entanglement reaches its maximum value. To show the point mentioned, I/Q distribution is analyzed for two different bias points. Fig. 2c shows the simulated single-shot I/Q distributions for the readout outcomes when the flux bias is set to $\phi/\phi_0 \sim 0.70$, a detuned regime where the tunable qubit and interior qubits are weakly coupled. The blue and orange point clouds correspond respectively to the logical states |0> and |1> of the targeted qubit, each representing 1200 simulated measurement shots. The clusters are clearly separated along the In-phase axis, and their Gaussian-like spread stems from the imposed classical noise characterized by $\kappa_{SNR}$. The black dashed line denotes the optimal discrimination boundary obtained from projecting the I/Q data along the maximal-separation axis. The deterministic mean I/Q values ($\mu_0$, $\mu_1$) extracted from the Monte Carlo solver are marked with black and red crosses. The high degree of spatial separation indicates strong dispersive readout contrast and yields a separate fidelity of approximately 0.9929. In the same way and same condition, Fig 2.d illustrates the I/Q-plane statistics under the entangled operating point, $\phi/\phi_0 \sim 0.82$, corresponding to the avoided-crossing region observed in the spectroscopy. Here, the tunable qubit strongly hybridizes with an interior qubit, forming correlated dressed states that jointly influence the resonator response. Despite the entanglement, the measured I/Q distributions remain well separated, with slightly increased contrast between the two logical manifolds. The deterministic means ($\mu_0$, $\mu_1$) shift further apart compared to Fig. 2c, demonstrating that the collective qubit–resonator system produces a larger state-dependent phase shift of the readout tone. Consequently, the simulated separation fidelity rises to about 0.9950, indicating that the hybridized system not only supports coherent coupling but also enhances the readout contrast. This behavior verifies that the engineered multi-qubit Hamiltonian enables a controllable entanglement regime where information encoded in the coupled qubits can still be extracted with near-optimal fidelity. In practical terms, Fig. 2d highlights the operational sweet spot of the device: the region near the avoided crossing simultaneously maximizes entanglement strength and readout discrimination, thereby providing an efficient platform for scalable multi-qubit measurement protocols.

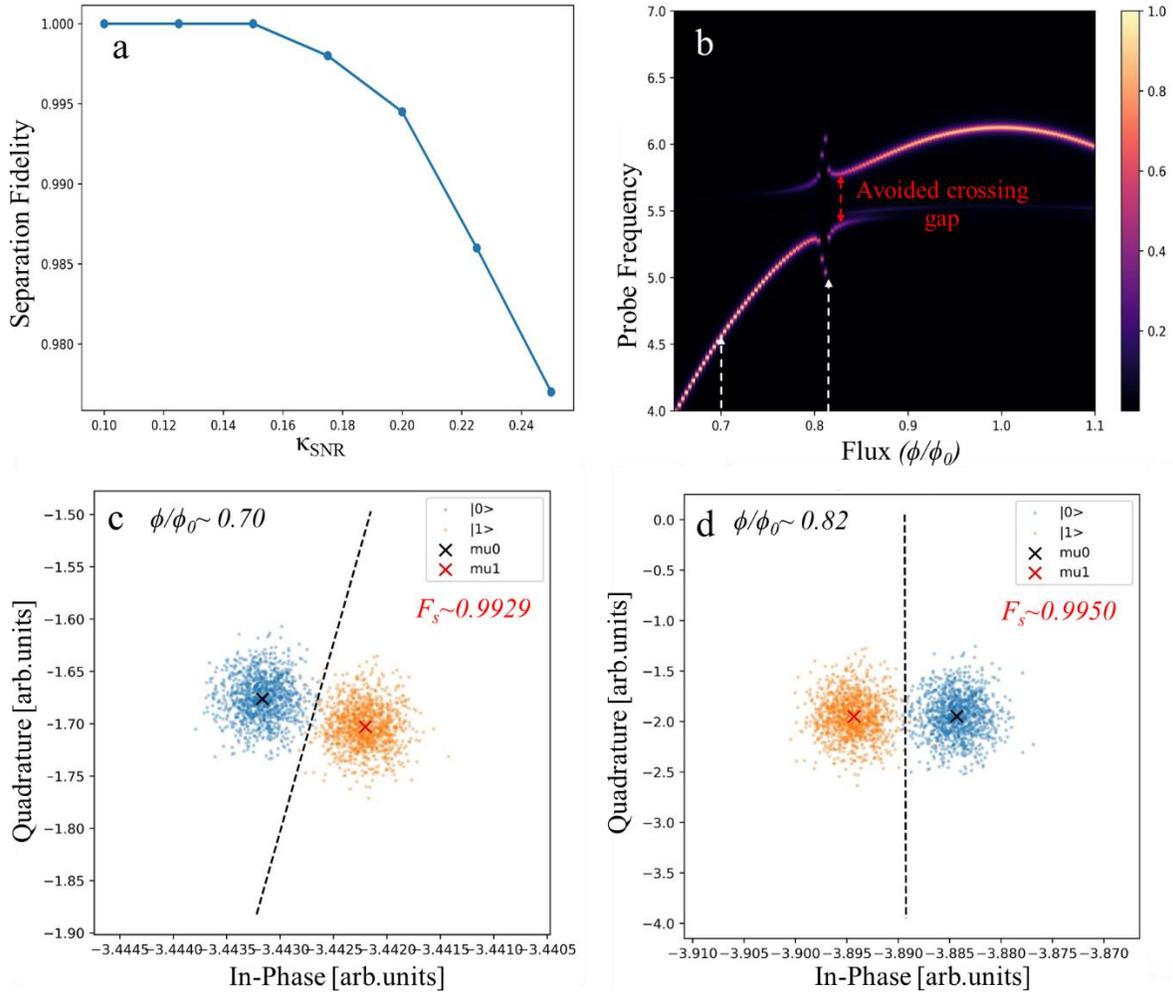

Fig. 2 Flux-tunable entanglement and readout fidelity in the coupled multi-qubit system.
a) Separation fidelity as a function of the signal-to-noise ratio degradation factor $\kappa_{SNR}$, showing that high-fidelity discrimination is preserved up to moderate noise levels before rapidly decreasing beyond $\kappa_{SNR} >$ 0.2. b) Two-tone spectroscopy of the tunable qubit as a function of normalized magnetic flux $\phi/\phi_0$, revealing an avoided crossing near $\phi/\phi_0 \sim 0.82$ that indicates coherent hybridization between the tunable and interior qubits. c) and d) I/Q-plane distributions of the measured quadratures respectively at $\phi/\phi_0 \sim$ 0.70 and $\phi/\phi_0 \sim 0.82$. The detuned case (c) shows clearly separated Gaussian clusters corresponding to |0> and |1> states, while at the avoided crossing (d) the states become entangled yet retain maximal separation fidelity ($F_s \sim 0.9950$). These results demonstrate that flux-controlled coupling enables entanglement without degrading readout performance; $\kappa_{SNR} = 0.2$.

To further strengthen the applicability and universality of the proposed quantum analog circuit, it is essential to establish its quantum gate–based equivalent representation. While the total Hamiltonian accurately describes the continuous physical dynamics of the superconducting circuit, expressing it in the Pauli-operator form enables its implementation within quantum gate computing frameworks [22-26]. This transition from an analog to a gate-level model allows the derived interactions—such as tunable qubit–qubit coupling and cross-resonance effects—to be represented as unitary gate operations. Incorporating this equivalent mapping not only validates that the designed circuit can reproduce the same entanglement–fidelity behavior on a programmable quantum processor but also introduces a novel analog–gate circuit co-design perspective. Such an approach provides a unified platform where hardware-level Hamiltonian engineering and algorithmic gate synthesis converge, enabling optimization through quantum gate Hamiltonian learning. Thus, including the gate-level equivalent circuit extends the scope of this work beyond hardware realization, demonstrating that the proposed architecture can serve as a universal, reconfigurable quantum module. To generate the quantum gate version of the quantum circuit designed, it is necessary to construct the mapping Hamiltonian (expressed in terms of Pauli operators). However, it is necessary to point out that the mapping Hamiltonian and its corresponding quantum gate circuit provide a computationally tractable framework for AI-based training and optimization, unlike the full physical quantum circuit governed by complex bosonic Hamiltonians [22-26]. It is because real superconducting systems involve nonlinear couplings, dissipation, and noise sources that make gradient evaluation and parameter optimization extremely challenging. In contrast, the mapping Hamiltonian reduces the system to a symbolic and differentiable form compatible with quantum machine-learning and reinforcement-learning algorithms. The mapping Hamiltonian serves as a simplified yet physically equivalent representation of the total Hamiltonian of the quantum circuit, enabling the translation of complex analog dynamics into a qubit-based computational framework. In a realistic superconducting chip, the total Hamiltonian contains multiple bosonic degrees of freedom—such as resonator modes, transmon anharmonic oscillators, and their mutual couplings—making it analytically and numerically intractable for large-scale simulation. To bridge this complexity, the system is mapped into an effective Pauli-operator Hamiltonian, where each mode is represented as a two-level system. This transformation is achieved through standard quantization and truncation procedures, replacing ladder operators such as $b, b^+$ with Pauli matrices $\sigma_x, \sigma_y$, and $\sigma_z$. Furthermore, by applying the Schrieffer–Wolff transformation in the dispersive regime, the intermediate resonator degrees of freedom are adiabatically eliminated, resulting in effective exchange and cross-Kerr interactions between qubits [22, 26]. The resulting mapping Hamiltonian therefore captures all essential physics—such as qubit–qubit coupling, tunable hybridization, and dispersive readout shifts—while operating within a compact spin-based formalism that is compatible with quantum gate modeling. In this way, the mapping Hamiltonian acts as the intermediate layer between analog quantum hardware and quantum gate operations, providing a universal framework for designing, optimizing, and benchmarking quantum processors. Therefore, regarding the total Hamiltonian derived, the mapping Hamiltonian is presented as:

$$H_{\text{map}} = \frac{\hbar \omega_c(\Phi)}{2} \sigma_c^z + \sum_{j=1}^{4} \frac{\hbar \omega_j}{2} \sigma_j^z + \sum_{i=1}^{4} \frac{\hbar \omega_i}{2} \sigma_i^z + \sum_{j=1}^{4} \hbar \chi_j \sigma_j^z (a_j^\dagger a_j) + \sum_{i=1}^{4} \hbar \chi_i \sigma_i^z (b_i^\dagger b_i) + \sum_{j=1}^{4} \hbar J_{jc}(\sigma_j^- \sigma_c^+ + \sigma_j^+ \sigma_c^-) + \sum_{i,j} \hbar J_{ij}(\sigma_i^- \sigma_j^+ + \sigma_i^+ \sigma_j^-) + \hbar \omega_o a^\dagger a + \sum_{k=1}^{4} \hbar \chi_k \sigma_k^z (a^\dagger a) \quad (5)$$

where $J_{ij} = g_i*g_j*(1/\Delta_i + 1/\Delta_j)$ and $J_{jc} = g_c*g_j*(1/\Delta_c + 1/\Delta_j)$. By considering $a = \sigma^-$, $a^+ = \sigma^+$, and $a^+a = 0.5(I-\sigma_z)$, Eq. (5) can be reformulated to generate quantum gates through the decomposition of the unitary operator $U(\theta) = \exp[-H_{\text{map}}(\theta)]$. The operator $U(\theta)$ describes the evolution of the quantum state as the circuit parameter when $\theta$ is changed [22-28]. In this context, $\theta$ may represent any key parameter of the quantum system—such as the qubit frequency, qubit–resonator coupling strength, or qubit–qubit interaction. Notably, such parameters can be optimized using AI-driven techniques to enhance the overall performance and adaptability of the quantum circuit.

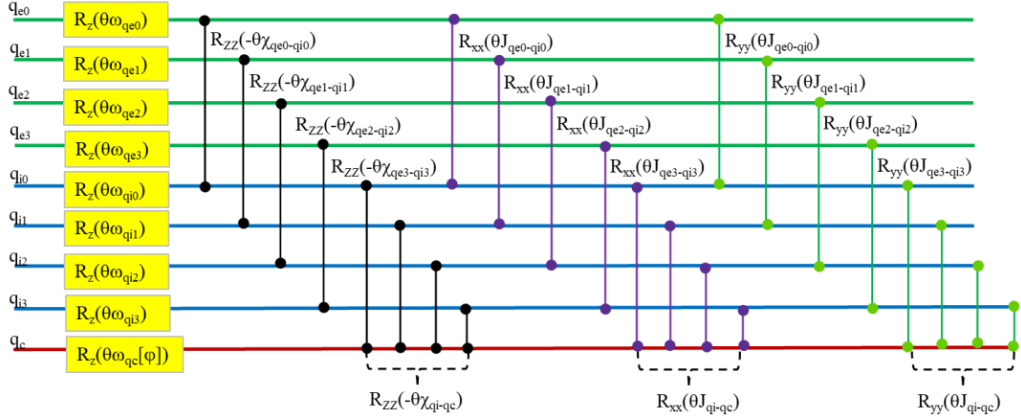

Fig. 3 Quantum gate circuit derived from the mapping Hamiltonian of the proposed superconducting quantum chip. Each qubit line represents an exterior, interior, or tunable qubit, and the $R_z$, $R_{xx}$, and $R_{yy}$ gates correspond respectively to dispersive, exchange, and hybrid interactions; in this circuit for simplicity the exterior qubits connotated with subscript "$q_e$" and the interior ones are indicated with "$q_i$".

Thus, the quantum gate that comprises the quantum analog circuit can be made using the unitary operator $U(\theta)$. Eventually using $U(\theta)$, the quantum gate circuit established is shown in Fig. 3, which represents the mapped equivalent of the superconducting analog chip architecture. Each qubit line corresponds to either an exterior, interior, or tunable qubit, while the vertical two-qubit gates reflect the effective interactions derived from the system's total Hamiltonian. The sequence of $R_z$, $R_{xx}$, and $R_{yy}$ operations correspond to the dispersive, exchange, and hybrid coupling terms ($\chi$, $J$) obtained via the Schrieffer–Wolff transformation in the mapping Hamiltonian. Specifically, the $R_{zz}$ gates represent dispersive shifts responsible for readout fidelity control through the exterior qubits, whereas the $R_{xx}$ and $R_{yy}$ gates, specifically between the interior qubits and tunable qubit, simulate coherent exchange processes that generate and sustain entanglement among the interior qubits and the central tunable qubit. The circuit is designed symmetrically so that the entanglement pathways and fidelity-enhancing couplings coexist without destructive interference. Consequently, this quantum gate model effectively emulates the complete quantum dynamics of the proposed hardware, allowing computational exploration, AI-assisted optimization, and benchmarking of entanglement–fidelity co-optimization strategies without requiring a physical device. The quantum gate circuit represents an abstraction of the physical superconducting chip, derived from the total Hamiltonian through the mapping Hamiltonian. While it accurately reproduces the logical operations and quantum dynamics, it does not directly reveal the physical mechanisms such as how specific couplings or resonators generate and preserve entanglement and fidelity. From the gate structure, one can infer which qubits interact or become entangled (via $R_{xx}/R_{yy}$ gates) and which contribute to dispersive readout or fidelity enhancement (via $R_{zz}$ gates). However, the physical realization of these interactions, defined by tunable flux qubit, resonator detuning, and circuit topology, remains hidden within the underlying analog model. The relationship between the physical circuit and the gate circuit is therefore not strictly reversible. The total Hamiltonian uniquely determines the gate model, but multiple hardware configurations can implement the same gate sequence. Nevertheless, by analyzing the coupling topology and gate dependencies, one can approximate a physical design consistent with the gate model. In the following, the QGHNN [22-24] is applied on the mapping Hamiltonian expressed in Eq. 3 to optimize the separate fidelity. Evaluating the fidelity directly from the full stochastic simulation is computationally expensive; each fidelity estimate requires evolving two quantum states under a 9-qubit Hamiltonian with multiple collapse operators over many trajectories. Searching optimization space directly with such simulations is therefore infeasible. To overcome this issue, we use QGHNN in which the code trains the mapping Hamiltonian expressed in Eq. 3 using Neural Network [22, 23]. The approach is schematically completely illustrated in Fig. 4.

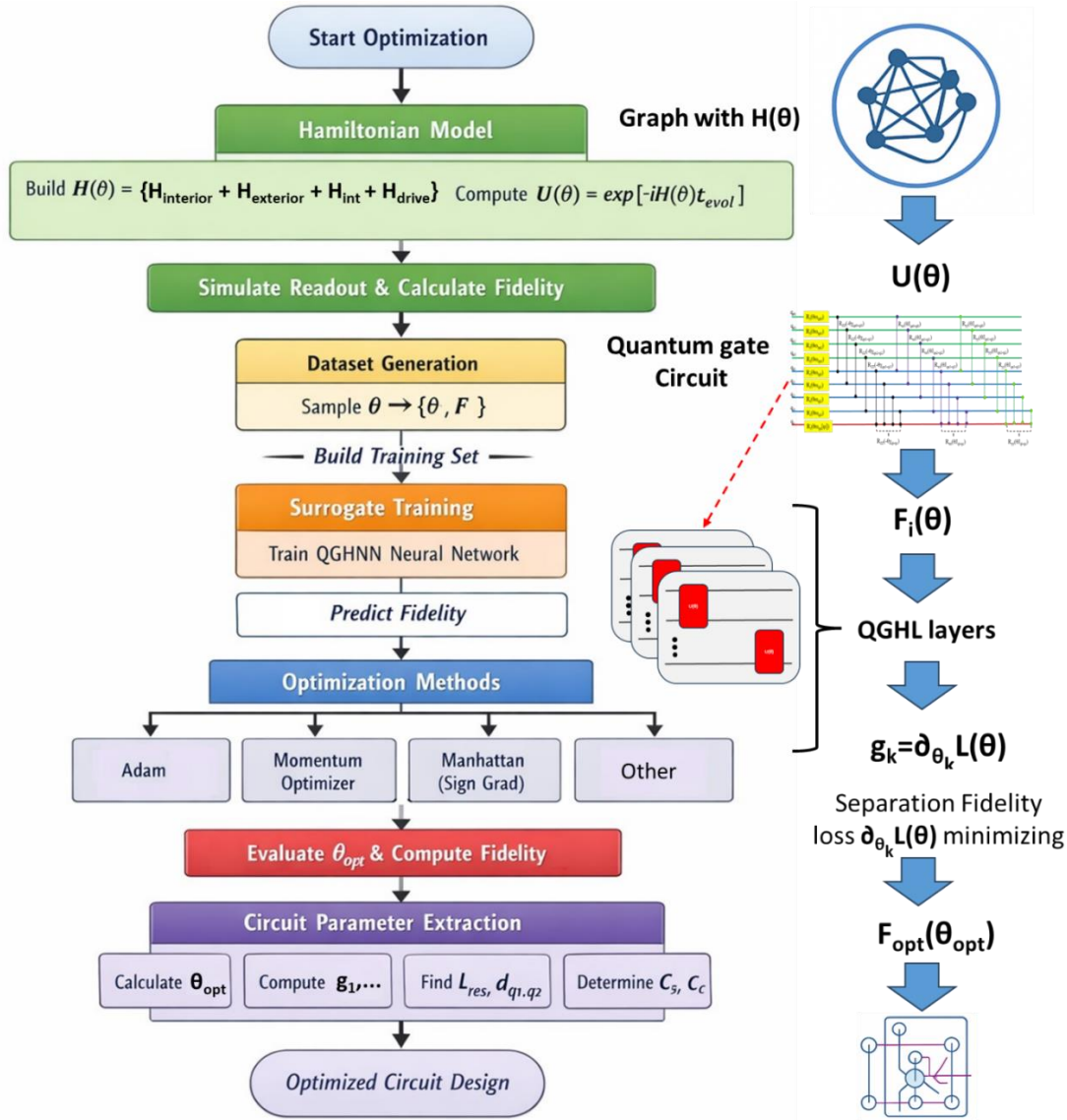

Fig. Flowchart of the physics-informed quantum circuit optimization pipeline, from Hamiltonian modeling and fidelity evaluation to quantum graph Hamiltonian learning (QGHL) and training, gradient-based optimization, and extraction of circuit-level design parameters.

The diagram illustrates the complete workflow of physics-aware optimization for the quantum circuit discussed. The process starts from a parameterized Hamiltonian $H(\theta)$, incorporating qubits, resonators, all interaction terms, and drive term. From this Hamiltonian, the unitary evolution operator $U(\theta)$ is computed and used to simulate the quantum readout dynamics. Measurement outcomes in the I/Q plane are processed to evaluate the separation fidelity, which quantifies readout distinguishability. These simulations are repeated to generate a dataset $\{\theta, F(\theta)\}$, forming the basis for training a QGHNN-based surrogate model that approximates the fidelity landscape. Optimization can then be performed using different strategies—

Adam [27], Momentum [28], Manhattan (sign-gradient) [29], or other methods—either directly on the physical model or via the surrogate. The optimized parameters $\theta_{opt}$ are finally translated into circuit-level quantities such as coupling rates, resonator length, capacitances, and Purcell decay rates, enabling a concrete, fabrication-ready quantum circuit design. The QGHNN is implemented as a physics-informed model that learns the nonlinear mapping between the device parameter vector θ (interior qubit frequencies, exterior qubit frequencies, resonator frequencies, coupling strengths, and the tunable flux parameter) and the resulting state-separate fidelity obtained from full master-equation simulations. The QGHNN receives vector θ and processes it through a sequence of fully connected layers. The model is trained on a dataset consisting of many pairs (θ, $F_s$), where $F_s$ denotes the single-shot state-separate fidelity computed from stochastic quantum trajectories. The QGHNN therefore learns a smooth functional approximation of the forward map (θ→$H_{map}$(θ)→ρ(t)→$F_s$) without explicitly reconstructing the intermediate Hamiltonian or solving the Lindblad equation during inference. Once trained, the network provides extremely fast fidelity predictions, enabling efficient optimization of circuit parameters in a high-dimensional landscape. In the implemented workflow, the optimizer repeatedly queries the QGHNN to estimate fidelity for candidate parameter sets. The simulation workflow is built to evaluate and optimize the separate fidelity of a tunable qubit embedded in a small multi-qubit architecture with integrated dispersive elements. The model includes (i) individual qubit energies, (ii) tunable-flux–dependent transmon frequency modulation through $E_J(\phi)$, (iii) effective exchange interactions between each interior/exterior qubit and the tunable qubit, (iv) dispersive χ-type ZZ interactions arising from integrated-out exterior qubits, and (v) a weak readout-drive term applied to the candidate qubit.

To calculate I/Q (Eq. 3 and Eq. 4), the time evolution under this Hamiltonian is computed via *mcsolve* [18], allowing stochastic quantum trajectories to emulate decay and dephasing. For each candidate parameter set $\theta$, the code prepares |0> and |1>and states on the qubit measured, evolves them, and then extracts approximate I/Q readout quantities by integrating expectation values of a readout operator. These raw values are used to construct synthetic measurement clouds, project the distributions along the optimal discrimination axis, fit approximate Gaussians, and finally compute the single-shot separate fidelity. The results of the simulation, before and after of the OGHNN are presented in Fig. 5. To ensure consistency with Fig. 2, this simulation was performed using a fixed SNR, with $\kappa_{SNR}$ = 0.2, while sweeping $\phi/\phi_0$ from 0.6 to 1.1. Additionally, the number of shots was reduced from 1500 to 800 to shorten the simulation time. In Fig. 5a, for $\phi/\phi_0$ = 0.6, the simulated I/Q measurement outcomes corresponding to the qubit states |0> and |1> form two slightly overlapping clusters. The intersection between the blue and orange point clouds indicates that the readout operator produces semi-indistinguishable trajectories under the unoptimized Hamiltonian parameters. As a result, the projected one-dimensional histograms show an overlap, reflecting a slight poor but fair separate fidelity. In contrast, Fig. 5b demonstrates the effect of using the QGHNN to guide parameter optimization. The optimized parameter vector yields two well-separated I/Q clusters with noticeably reduced spread and minimal overlap. This enhancement arises because QGHNN identifies parameter combinations—qubit frequencies, coupling strengths, and flux bias—that maximize the dynamical contrast between the |0> and |1> trajectories during measurement. The data generated before and after QGHNN optimization are listed in Table. 1. Together, the figures provide clear visual evidence that the QGHNN-guided design approach successfully reshapes the device Hamiltonian's variables to achieve high-fidelity quantum-state separation.

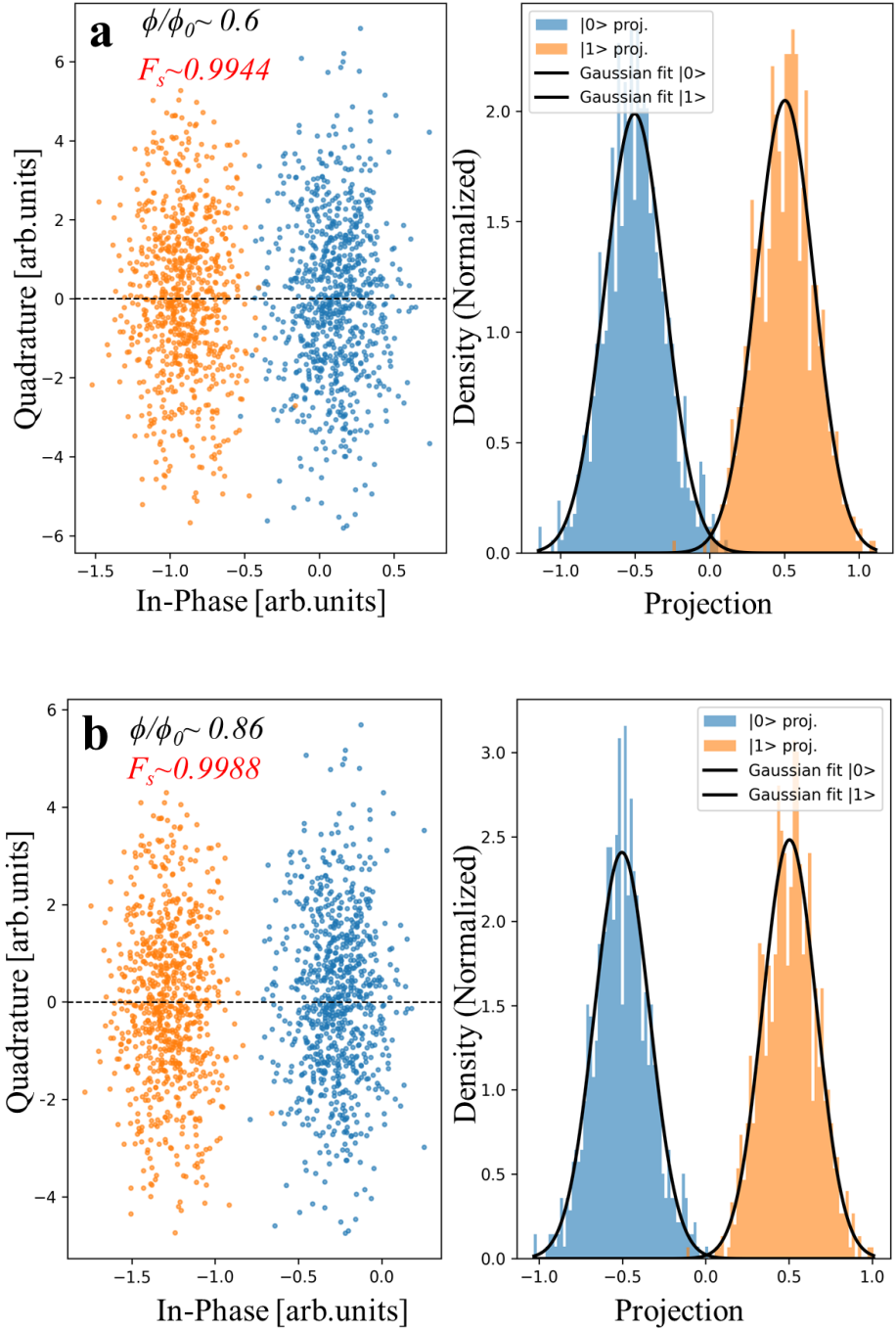

Fig. 4 I/Q measurement distributions a) before QGHNN optimization for $\phi/\phi_0 = 0.6$, b) after QGHNN optimization for $\phi/\phi_0 = 0.86$; Scatter plot of the simulated single-shot measurement outcomes corresponding to the qubit states |0⟩ (blue) and |1⟩ (orange) prior to optimization; $\kappa_{SNR} = 0.2$.

However, the aim of this work, in the future, will be to use of optimized parameters in professional quantum-circuit design (some of important parameters to design an optimized quantum circuit are listed in Table. 2). The optimized parameter vector returned by QGHNN, denoted $\theta_{opt}$, contains explicit physical quantities directly usable in hardware design and can be translated into other circuit-level parameters capacitances, resonators length, the distance between two adjacent qubits and others. For example, the optimized interior qubit frequencies ($\omega_j^{(int)}$) and exterior qubit frequencies $\omega_i^{(ext)}$ can be mapped to Josephson energies and capacitor layouts to manipulate the quantum circuit topology. The resonator frequencies $\omega_r$ determine resonator lengths and characteristic impedances. The coupling constants $\{\chi_i, \chi_j, \chi_k, g_i, g_j, g_c\}$ addressed in mapping Hamiltonian introduced in Eq. 5 define target capacitive or inductive coupling strengths between nodes of the circuit, guiding suitable placement and spacing. Finally, the optimized flux parameter $\phi/\phi_0$ specifies the bias point at which the tunable qubit operates, which informs the flux-bias control circuitry and the design of the SQUID loop asymmetry. In practice, one incorporates $\theta_{opt}$ into the engineering workflow by (i) translating each parameter into circuit-element values, (ii) conducting electromagnetic simulations (e.g., HFSS, COMSOL) to validate and fine-tune the target frequencies, and (iii) after validating, embedding the resulting chip-level parameters into a control-electronics pipeline to finalize the quantum circuit design. Because the optimized parameters result in substantially improved discrimination fidelity, the final device is expected to achieve higher fidelity and lower measurement error. Thus, the QGHNN-guided optimization does not merely improve simulated fidelity; it provides a concrete set of physically interpretable design parameters that can be transferred directly into professional quantum-circuit fabrication procedures.

Table 1. Parameters before and after QGHNN optimization.

| | Before QGHNN optimization | After QGHNN optimization ($\theta_{opt}$) |
|---|---|---|
| $\omega_j^{(int)}/2\pi$ (GHz) | 5.52, 5.53, 5.57, 5.58 | 5.56, 5.55, 5.53, 5.55 |
| $\omega_i^{(ext)}/2\pi$ (GHz) | 5.2, 5.4, 5.6, 5.8 | 5.24, 5.42, 5.43, 5.64 |
| $\omega_r/2\pi$ (GHz) | 5.10, 5.30, 5.75, 5.90 | 5.38, 5.87, 5.81, 5.22 |
| $\chi_i$ (Mrad/s) | 62.83, 62.83, 62.83, 62.83 | 700, 421, -346, 118 |
| $\chi_j$ (Mrad/s) | 62.83, 62.83, 62.83, 62.83 | -38.5, 79, 20.7, 364 |
| $\chi_k$ (Mrad/s) | 125.66, 125.66, 125.66, 125.66 | 435, -290, 141, 225 |
| $g_i$ (Mrad/s) | 62.83, 62.83, 62.83, 62.83 | 528, -19.61, 517, 422 |
| $g_j$ (Mrad/s) | 62.83, 62.83, 62.83, 62.83 | 662, 340, -4.07, 4.88 |
| $g_c$ (Mrad/s) | 62.83 | 92.58 |
| $\phi/\phi_0$ | 0.6 | 0.86 |

Table 2. Quantum-circuit design parameters after QGHNN optimization ($Z_0$ = 50, $\alpha/2\pi$ [qubit-anharmonicity] ~ 200 MHz, $\varepsilon$ [effective dielectric constant] = 5.5).

| Parameter | Value | Definition |
|---|---|---|
| $L_{ri}$ (mm) | 7.19, 8.08, 7.09, 8.23 | Physical length of resonator |
| $\kappa_i/2\pi$ (MHz) | 3.35, 3.69, 3.65, 3.25 | Energy decay rate |
| $\Gamma_{Pi}/2\pi$ (KHz) | 2.37, 1.48, 1.79, 1.01 | Purcell decay rate |
| $C_\Sigma$ (fF) | 96.85 | Total qubit capacitance |
| $C_{ci}$ (fF) | 23.33, 21.18, 21.38, 24.01 | Coupling capacitance |
| $L_{di}$ (mm) | 3.55, 3.99 3.48, 4.1 | Exterior qubit-Interior qubit separation on chip |

Finally, the performance of the optimized quantum chip design is assessed using Grover's search algorithm [30,31], which provides a quadratic speedup for locating a marked item in an unstructured database of $N$ elements, requiring $O(\sqrt{N})$ operation. The step by step of the Grover algorithm is as follows: 1) prepare the initial state $|0\rangle^{\otimes n}$; 2) apply Hadamard $H^{\otimes n}$ to create superposition; 3) Apply the Oracle operator to mark the target O|x> = -|x>; 4) apply the Grover diffusion operator to amplify the probability of the target finding; 5) reaper the stages 3) and 4) for about √N; 6) Measurement. Since the quantum chip architecture considered in this work incorporates multiple qubit configurations, several representative 2-qubit pairs are selected to evaluate the implementation of Grover's algorithm. Nevertheless, regardless of the specific two-qubit pair under consideration, the underlying 2-qubit system is governed by the same total quantum Hamiltonian, which can be expressed as:

$$H_{eff} = \sum_{i=1}^{2} \frac{\hbar \tilde{\omega}_i}{2} \sigma_z^{(i)} + \frac{\hbar J_{12}}{2}\left(\sigma_x^{(1)}\sigma_x^{(2)} + \sigma_y^{(1)}\sigma_y^{(2)}\right) + \hbar \xi \sigma_z^{(1)}\sigma_z^{(2)} \quad (6)$$

where $\tilde{\omega}_i$ is the dressed qubit frequencies, $J_{12}$ is the transverse exchange coupling, and $\zeta$ is the effective ZZ coupling induced by the bus. The quantum gate circuit for 2-qubit Grover algorithm to find |11> is illustrated in fig. 5. It simply shows the successive operators that should be applied to successfully run a Grover algorithm.

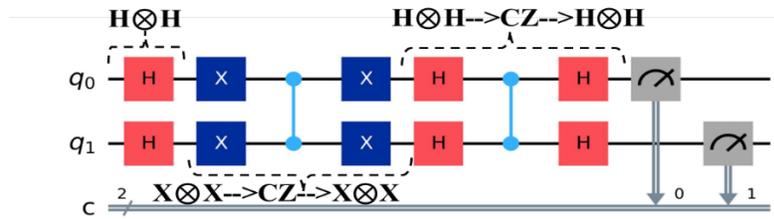

Fig. 5 Circuit for 2-qubit Grover algorithm to find |11>; initiating with Hadamard operator, applying Oracle operator, applying Grover diffusion operator, and finally measurement.

Table 3. 2-qubit Grover algorithm with different qubits combinations on Nine-qubit quantum chip (2048 shots); assumption for decoherence constants: $T_1 \sim 80$ μs, $T_2 \sim 70$ μs, $\kappa \sim 0.1/$μs.

| Qubits used | effective ZZ coupling ξ (MHz) | Accuracy |
|---|---|---|
| $Q_{11} \rightarrow \omega_{1*} \rightarrow Q_{22}$ | 0.2556 | 71.19% |
| $Q_{11} \rightarrow \omega_{11} \rightarrow Q_c$ | 0.7097 | 78.08% |
| $Q_1 \rightarrow \omega_1 \rightarrow Q_{11}$ | 0.2639 | 71.48% |
| $Q_1 \rightarrow \omega_{1*} \rightarrow Q_2$ | 0.5322 | 76.76% |

Table 3. 2-qubit Grover algorithm with a typical 2-qubit combination (int1_res_tun) on Nine-qubit quantum chip with different shots number.

| Number of shots | Accuracy |
|---|---|
| 512 | 77.93% |
| 1024 | 78.03% |
| 2048 | 78.08% |
| 4096 | 78.10% |

The simulation outcomes of the target state (assumed to be |11>) under representative decoherence parameters are summarized in Tables 3 and 4. As shown in Table 3, it is shown that the different 2-qubit configuration shows the different accuracy; this is contributed to the quantum circuit Hamiltonian and its coupling strength. For each pair of the qubit listed, the optimized parameters derived from the QGHNN ($\omega_1$, $\omega_2$, $\omega_r$, $\chi_1$, $\chi_2$, $J_{12}$) is used. Additionally, it is found that the Grover accuracy exhibits a strong dependence on the parameter ξ, which governs the Grover diffusion operator. Therefore, the effective ZZ interaction plays a more dominant role than the transverse coupling between the two qubits. Consequently, the qubit–

bus–mediated coupling, which controls the strength of the effective ZZ interaction, has a decisive impact on the performance of the Grover algorithm and can dramatically alter the success probability. These effects are discussed in detail in the following sections. Moreover, the accuracies achieved for different two-qubit configurations are comparable to those reported for the IBM Q 5 Tenerife quantum processor [30]. In addition, for a representative two-qubit system, the influence of the number of measurement shots on the Grover algorithm accuracy is investigated, and the corresponding results are presented in Table 4. Consistent with the findings in [30], the accuracy saturates for larger numbers of shots, showing only marginal variation beyond a certain threshold. In the following, we attempt to show that the effective ZZ interaction plays a more dominant role than the transverse coupling between the two qubits to enhance the Grover accuracy.

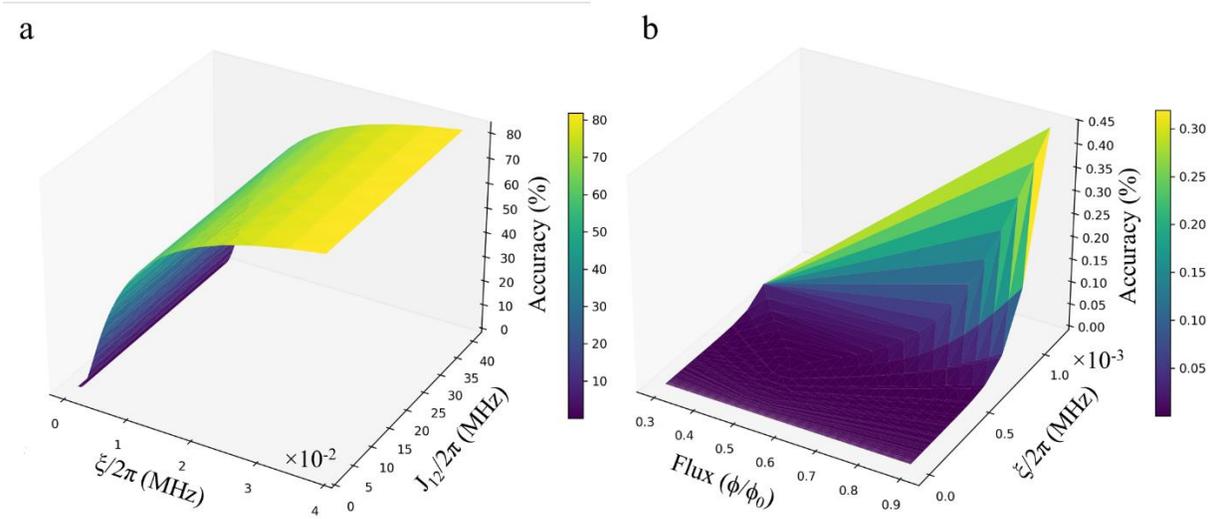

Fig. 6 a) ZZ and transvers coupling effect on the Grover accuracy (%); b) ZZ coupling and flux φ effect on the Grover accuracy (%).

Fig. 6a illustrates the combined influence of the effective ZZ coupling strength $\xi$ and the transverse exchange interaction $J_{12}$ on the Grover algorithm accuracy. The results clearly show that the performance is predominantly governed by the ZZ interaction: as $\xi$ increases, the success probability rises rapidly and eventually saturates at high accuracy values. In contrast, variations in the transverse coupling $J_{12}$ have a comparatively weaker impact, mainly producing minor modulations once a sufficient ZZ interaction is established. This behavior confirms that the phase accumulation required for the Grover diffusion operator is primarily enabled by the effective ZZ coupling, while the transverse interaction plays a secondary role in fine-tuning the dynamics. Consequently, engineering a strong and controllable ZZ interaction is crucial for achieving high-fidelity Grover operations in bus-mediated two-qubit architectures. The other interesting task studied in this work is to consider the effect of flux $\phi$ on the Grover accuracy. From Fig. 2, it was shown that near $\phi/\phi_0 \sim 0.82$ the avoided crossing is occurred and also the separate fidelity becomes maximum. In line with, Fig. 6b presents the dependence of the Grover accuracy on the external flux bias $\phi$ as well as the effective ZZ coupling $\xi$. For small values of $\xi$, the accuracy remains close to zero across the entire flux range, indicating that flux tuning alone is insufficient to enhance the algorithm performance without an adequate ZZ interaction. However, as $\xi$ increases, a pronounced enhancement in accuracy is observed, particularly close to $\phi/\phi_0 \sim 0.8$ flux values, where the avoided crossing occurred and also the interior qubits create the entanglement (see Fig. 2). Thus, entanglement enhancement, high-fidelity operation, and improved Grover accuracy converge at a common operating point, indicating that the QGHNN-optimized quantum chip effectively resolves the trade-off—particularly that induced by unwanted ZZ interactions—in conventional quantum architectures. Indeed, this behavior reflects the flux-controlled modification of the qubit frequencies and qubit–bus detunings, which in turn regulate the effective ZZ coupling strength. The strong correlation between $\phi$, $\xi$, and the algorithmic success probability demonstrates that flux tunability provides an efficient factor to optimize the entanglement, high fidelity, and also Grover diffusion process through indirect control of the ZZ interaction.

**Conclusions:**

In this work, we have presented a comprehensive framework for the design, optimization, and algorithmic validation of a scalable superconducting quantum-chip architecture that overcomes a long-standing trade-off between strong entanglement, high operational fidelity, and Grover algorithm accuracy. By introducing a hybrid nine-qubit architecture composed of interior and exterior transmon qubits interconnected via a flux-tunable qubit and a distributed resonator network, we demonstrated that dynamic control of qubit–qubit interactions can be achieved without sacrificing readout performance. The interior qubits and tunable coupler form a controllable entanglement core, while the exterior qubits operate deep in the dispersive regime, enabling high-fidelity state separation under realistic noise conditions. Through full Hamiltonian modeling and Lindblad master-equation simulations, it is shown that the proposed architecture supports robust entanglement formation characterized by engineered avoided crossings, while maintaining separation fidelity exceeding 0.995. Building upon this physical platform, a mapping-Hamiltonian is developed which is enabled a gate-level abstraction of the circuit dynamics and facilitated the application of QGHNN. By learning the nonlinear relationship between device parameters and single-shot separation fidelity, the QGHNN enabled efficient optimization in a high-dimensional design space, yielding an optimized parameter set that is physically interpretable and directly transferable to hardware fabrication workflows. The optimized architecture achieved separation fidelity approaching 0.9988, confirming the effectiveness of the physics-informed learning approach. Notably, the optimized parameters are physically meaningful and immediately transferable to real hardware design. The updated qubit and resonator frequencies map directly to Josephson energies, capacitor geometries, and resonator lengths; the optimized coupling constants define target capacitive and inductive interaction strengths; and the refined flux-bias point determines the operating regime of the tunable qubit. These optimized values can be integrated into professional quantum-circuit engineering pipelines, including electromagnetic simulation (HFSS, COMSOL), layout verification, and calibration planning. Finally, to assess the computational capability of the optimized quantum chip, Grover's search algorithm is implemented on representative two-qubit subsystems consisting of different configurations. The results revealed that the Grover algorithm accuracy is predominantly governed by the effective ZZ interaction mediated by the bus resonator, while the transverse exchange coupling plays a secondary role. Systematic studies further showed that flux tunability provides a powerful control knob to simultaneously enhance entanglement strength, suppress detrimental crosstalk, and optimize the Grover diffusion process. As a result, the optimized architecture achieves Grover accuracies comparable to those reported for the IBM Q5 Tenerife processor, demonstrating competitive algorithm-level performance. Eventually, this study establishes a unified and experimentally relevant strategy in which architectural design, Hamiltonian-level modeling, machine-learning-based optimization, and algorithmic benchmarking are seamlessly integrated. The results demonstrate that entanglement, high fidelity, and high Grover accuracy can be co-optimized within a single reconfigurable superconducting platform, effectively resolving the limitations imposed by static ZZ crosstalk in conventional designs. Finally, the proposed QGHNN-guided approach provides a scalable pathway toward the systematic engineering of next-generation superconducting quantum processors capable of reliable, high-performance quantum computation.

**References:**
1. Xue, X., Patra, B., van Dijk, J.P.G. et al. CMOS-based cryogenic control of silicon quantum circuits. Nature 593, 205–210 (2021). https://doi.org/10.1038/s41586-021-03469-4.
2. B. Zhao *et al.*, EDA-Q: Electronic Design Automation for Superconducting Quantum Chip, *arXiv:2502.15386*, 2025. https://doi.org/10.48550/arXiv.2502.15386.
3. P. D. Sawant *et al.*, Quantum Dots Based Quantum Chips and Their Applications in Quantum Analytics, Software Development, Scalable Cloud and Mobile-Edge Solutions, *Journal of Future Internet and Hyperconnectivity,* 2, (2025).
4. A. D. G. Oquendo, A. Nadir, T. Jonuzi, S. Patra, N. Kanti Sinha, R. Orús, S. Mugel, Accelerating Photonic Integrated Circuit Design: Traditional, ML and Quantum Methods, arXiv:2506.18435 [quant-ph], 2025. https://doi.org/10.48550/arXiv.2506.18435.
5. A Salmanogli, H Zandi, S Hajihosseini, M Esmaeili, MH Eskandari, M. Akbari, Purcell Rate Suppressing in a Novel Design of Qubit Readout Circuit, arXiv e-prints, arXiv: 2411.07153, 2025.
6. A Salmanogli, H Zandi, MET Poshti, MH Eskandari, E Zencir, M Akbari, Design of Advanced Readout and System-on-Chip Analog Circuits for Quantum Chip, arXiv preprint arXiv:2506.16361, 2025.
7. A Salmanogli, A Bermak, Design of Fully Integrated 45 nm CMOS System-on-Chip Receiver for Readout of Transmon Qubit, AVS Quantum Sci. 7, 042003 (2025). https://doi.org/10.1116/5.0291986
8. Majer, J., Chow, J., Gambetta, J. *et al.* Coupling superconducting qubits via a cavity bus. *Nature* 449, 443–447 (2007). https://doi.org/10.1038/nature06184.
9. M. O. Scully and M. S. Zubairy, Quantum Optics (Cambridge University Press, Cambridge, 1997).
10. D. J. van Woerkom, P. Scarlino, J. H. Ungerer, C. M¨uller, J. V. Koski, A. J. Landig, C. Reichl, W. Wegscheider, T. Ihn, K. Ensslin, and A. Wallraff, Microwave photon-mediated interactions between semiconductor qubits, Phys. Rev. X 8, 041018, 2018. https://doi.org/10.1103/PhysRevX.8.041018.
11. Peng Zhao, Yingshan Zhang, Guangming Xue, Yirong Jin, and Haifeng Yu, Tunable coupling of widely separated superconducting qubits: A possible application towards a modular quantum device,
Appl. Phys. Lett. 121, 032601 (2022). https://doi.org/10.1063/5.0097521.
12. Pranav Mundada, Gengyan Zhang, Thomas Hazard, and Andrew Houck, Suppression of Qubit Crosstalk in a Tunable Coupling Superconducting Circuit, Phys. Rev. Applied 12, 054023 (2019). https://doi.org/10.1103/PhysRevApplied.12.054023.
13. Fei Yan, Philip Krantz, Youngkyu Sung, Morten Kjaergaard, Daniel L. Campbell, Terry P. Orlando, 1 Simon Gustavsson, and William D. Oliver, Tunable Coupling Scheme for Implementing High-Fidelity Two-Qubit Gates, Phys. Rev. Applied 10, 054062 (2018).
https://doi.org/10.1103/PhysRevApplied.10.054062.
14. X. Li, Y. Sung, S. Chakram, et al., "Suppression of Qubit Crosstalk in a Tunable Coupling Superconducting Circuit," Phys. Rev. Applied 12, 054023, 2019.
https://doi.org/10.1103/PhysRevApplied.12.054023.
15. A. Salmanogli and H. S. Geçim, "Optical and Microcavity Modes Entanglement by Means of Plasmonic Opto-Mechanical System," IEEE Journal of Selected Topics in Quantum Electronics, vol. 26, pp. 1-10, 2020. doi: 10.1109/JSTQE.2020.2987171.

16. A. Salmanogli, D. Gökçen, and H. S. Geçim, Entanglement of Optical and Microcavity Modes by Means of an Optoelectronic System, Phys. Rev. Appl. 11, 024075 (2019). https://doi.org/10.1103/PhysRevApplied.11.024075.

17. A. Salmanogli, Entangled microwave photons generation using cryogenic low noise amplifier (transistor nonlinearity effects), Quantum Sci. Technol. 7, 045026 (2022). 10.1088/2058-9565/ac8bf0.

18. J. R. Johansson, P. D. Nation, and F. Nori, QuTiP 2: A Python framework for the dynamics of open quantum systems, Comput. Phys. Commun. 184, 1234 (2013). https://doi.org/10.1016/j.cpc.2012.11.019

19. A. Salmanogli, H. Zandi and M. Akbari, "Blochnium-Based Josephson Junction Parametric Amplifiers: Superior Tunability and Linearity," IEEE Journal of Selected Topics in Quantum Electronics, vol. 31, Quantum Materials and Quantum Devices, pp. 1-8, 2025. 10.1109/JSTQE.2024.3522509.

20. E. Jeffrey et al., Fast Accurate State Measurement with Superconducting Qubits, Phys. Rev. Lett. 112, 190504 (2014). DOI: https://doi.org/10.1103/PhysRevLett.112.190504

21. T. Walter, P. Kurpiers, S. Gasparinetti, P. Magnard, A. Potočnik, Y. Salathé, M. Pechal, M. Mondal, M. Oppliger, C. Eichler, and A. Wallraff, Rapid High-Fidelity Single-Shot Dispersive Readout of Superconducting Qubits, Phys. Rev. Applied 7, 054020 – Published 26 May, 2017. https://doi.org/10.1103/PhysRevApplied.7.054020.

22. W. Wang, QGHNN: A quantum graph Hamiltonian neural network, arXiv:2501.07986v1, https://doi.org/10.48550/arXiv.2501.07986, 2025.

23. J. Shi, W. Wang, X. Lou, S. Zhang, and X. Li, "Parameterized hamiltonian learning with quantum circuit," IEEE Transactions on Pattern Analysis and Machine Intelligence, vol. 45, no. 5, pp. 6086–6095, 2022.

24. J. Y. Araz and M. Spannowsky, "Quantum-probabilistic Hamiltonian learning for generative modeling and anomaly detection," Physical Review A, vol. 108, no. 6, p. 062422, 2023.

25. H.-Y. Huang, M. Broughton, J. Cotler, S. Chen, J. Li, M. Mohseni, H. Neven, R. Babbush, R. Kueng, J. Preskill et al., "Quantum advantage in learning from experiments," Science, vol. 376, no. 6598, pp. 1182–1186, 2022.

26. J. Shi, T. Chen, W. Lai, S. Zhang and X. Li, "Pretrained Quantum-Inspired Deep Neural Network for Natural Language Processing," IEEE Transactions on Cybernetics, vol. 54, pp. 5973-5985, Oct. 2024, doi: 10.1109/TCYB.2024.3398692.

27. D. P. Kingma, J. Lei Ba, ADAM: A METHOD FOR STOCHASTIC OPTIMIZATION, arXiv:1412.6980v9 [cs.LG] 30 Jan 2017.

28. I. Sutskever, J. Martens, George Dahl, G. Hinton, On the importance of initialization and momentum in deep learning, *Proceedings of the 30th International Conference on Machine Learning*, PMLR 28(3):1139-1147, 2013.

29. V. Leplat, S. Mayorga, R. Hildebrand, and A. Gasnikov, NORM-CONSTRAINED FLOWS AND SIGN-BASED OP-TIMIZATION: THEORY AND ALGORITHMS, arXiv:2508.18510v1, 2025.

30. A. Mandviwalla, K. Ohshiro and B. Ji, "Implementing Grover's Algorithm on the IBM Quantum Computers," *2018 IEEE International Conference on Big Data (Big Data)*, Seattle, WA, USA, 2018, pp. 2531-2537, doi: 10.1109/BigData.2018.8622457.

31. P. J. Coles, S. Eidenbenz, S. Pakin, A. Adedoyin, J. Ambrosiano, P. Anisimov, W. Casper, G. Chennupati, C. Coffrin, H. Djidjev et al, "Quantum algorithm implementations for beginners," *arXiv preprint arXiv:1804.03719*, 2018.